\documentclass[11pt]{article}
\usepackage{graphicx}
\usepackage{authblk}
\usepackage{wrapfig}
\usepackage{amsmath}
\usepackage{geometry} % See geometry.pdf to learn the layout options. There are lots.
\usepackage{typearea}
\typearea{10}
\newcommand{\dru}{day$^{-1}$kg$^{-1}$keV$^{-1}$ }
\newcommand{\kevee}{keV$_\mathrm{ee}$ }

\title{
{\Large
DARK MATTER SEARCH WITH HIGH PURITY NaI(Tl) SCINTILLATOR}
}
\author{\Large{K.~Fushimi$^{*}$}}
\affil{Department of Physics, Tokushima University, 2-1 Minami Josanajima-cho, Tokushima city, Tokushima , 770-8506, Japan}
\author{Y.~Kanemitsu}
\author[2]{K.~Kotera}
\affil{Graduate School of Integrated Arts and Sciences, Tokushima University,
1-1 Minami Josanajima-cho, Tokushima city, Tokushima , 770-8502, Japan}
\author{D.~Chernyak}
\affil{Department of Physics and Astronomy, University of Alabama, Tuscaloosa, Alabama 35487, USA}
\author{H.~Ejiri}
\affil{Research Center for Nuclear Physics, Osaka University, 10-1 Mihogaoka Ibaraki city, Osaka, 567-0042, Japan}
\author{K.~Hata}
\affil{Research Center for Neutrino Science, Tohoku University, 6-3 Aramaki Aza Aoba, Aobaku, Sendai city, 
Miyagi, 980-8578, Japan }
\author{R.~Hazama}
\affil{Department of Environmental Science and Technology, Osaka Sangyo University, 3-1-1 Nakagaito, 
Daito city, Osaka, 574-8530, Japan}
\author{T.~Iida}
\affil{Faculty of Pure and Applied Sciences, University of Tsukuba, 1-1-1 Tennoudai, Tsukuba city, 
Ibaraki, 305-8571, Japan}
\author[5]{H.~Ikeda}
\author{K.~Imagawa}
\affil{I.~S.~C. Lab.~, 5-15-24 Torikai Honmachi, Settsu city, Osaka, 566-0052, Japan}
\author[5,9]{K.~Inoue}
\affil{Kavli Institute for the Physics and Mathematics of the Universe (WPI),
5-1-5 Kashiwanoha, Kashiwa city, Chiba, 277-8583, Japan}
\author{H.~Ishiura}
\affil{Department of Physics, Graduate School of Science, Kobe University, 1-1 Rokkodai-cho, Nada-ku, Kobe city, Hyogo, 657-8501, Japan}
\author{H.~Ito}
\affil{Institute for Cosmic Ray Research, The University of Tokyo, 5-1-5 Kashiwanoha, Kashiwa city, 
Chiba, 277-8583, Japan}
\author{T.~Kisimoto}
\affil{Department of Physics, Osaka University, 1-1 Machikaneyama-cho,
Toyonaka city,  Osaka 560-0043, Japan}
\author[5,9]{M.~Koga}
\author{A.~Kozlov}
\affil{National Research Nuclear University ``MEPhI'' (Moscow Engineering Physics Institute), Moscow, 115409, Russia}
\author[9,14]{K.~Nakamura}
\affil{Butsuryo College of Osaka, 3-33 Ohtori Kitamachi, Nishi ward, Sakai city, Osaka, 
593-8328, Japan }
\author[1]{R.~Orito}
\author[4]{T.~Shima}
\author[9,11]{Y.~Takemoto}
\author[4]{S.~Umehara}
\author[2]{Y.~Urano}
\author[8]{K.~Yasuda}
\author[12]{S.~Yoshida}

\date{}
\normalsize
\begin{document}

\maketitle

\begin{abstract}
A dark matter search project needs and extremely low background radiation detector since the expected event rate of dark matter 
is less than a few events in one year in one tonne of the detector mass.
The authors developed a highly radiopure NaI(Tl) crystal to search for dark matter.
The best combination of the purification methods was developed, resulting $^{\mathrm{nat}}$K and $^{210}$Pb were
less than 20 ppb and 5.7 $\mu$Bq/kg, respectively.

The authors will construct a large volume detector system with high-purity NaI(Tl) crystals.
The design and the performance of the prototype detector module will be reported in this article. 
\end{abstract}

\section{Introduction}
\subsection{Search for weakly interacting massive particles}
The observation of cosmic microwave background (CMB) gives 
the dark matter density $\Omega_{\mathrm{DM}}h^{2}=0.120\pm 0.001$ and baryon density
$\Omega_{\mathrm{b}}h^{2}=0.0224\pm 0.0001$,
where $h=H_{0}/(100$ km s$^{-1}$ Mpc$^{-1}$) and Hubble constant $H_{0}=0.674\pm0.005$ km s$^{-1}$ Mpc$^{-1}$  \cite{Planck2018}.
The observation of microlensing by the dark matter cluster revealed its distribution in the universe \cite{Oguri2017, Honma2012, Honma2015}.

The dark matter also exists in our galaxy; 
the rotation curve of the galaxy gives the dark matter density in the vicinity of our solar system as
$\rho_{\odot} = 0.39\pm 0.09 \textrm{\ GeV\ cm}^{-3}$,
and the rotation speed as $238\sim240$ km s$^{-1}$ \cite{Sofue2020}.

Theoretical particle physics proposes various candidates of dark matter.
Two strong candidates for dark matter have been discussed; one is the weakly interacting massive particles (WIMPs), 
another is axion and axion-like particles (ALPs).
This paper will discuss WIMPs detection by a high sensitivity NaI(Tl) scintillator array.
WIMPs is a hypothetical particle that has been proposed by the beyond standard theory.
The expected mass ranges a several hundred MeV to several TeV.\@
It interacts with the target nucleus via weak interaction; the typical cross section is less than $10^{-40}$ cm$^{2}$.
The expected event rate in a radiation detector is less than
$10^{-1}$ kg$^{-1}$day$^{-1}$ 
because of the small cross section, small number density, and low speed of the WIMPs motion.

The signal of WIMPs in a radiation detector is mainly due to the elastic scattering 
between the WIMPs and nucleus.
The recoil energy of nucleus lies below 100 keV due to the low speed of WIMPs.
One needs to design a radiation detector with both low background and
low energy threshold.

One need to get a significant character to distinguish the WIMPs signal from the background since
the energy spectrum of recoil nucleus does not have a prominent structure.
The annual modulation of the energy spectrum is useful identifier for WIMPs.
The earth revolves around the sun with 31 km s$^{-1}$;
the orbital plane of the earth declines to the velocity of the solar system with 60$^{\circ}$.
The relative velocity of the WIMPs has a maximum at the beginning of June and a minimum at the beginning of  December \cite{Freese1988}.

\begin{wrapfigure}{r}{0.5\linewidth}
\includegraphics[width=\linewidth]{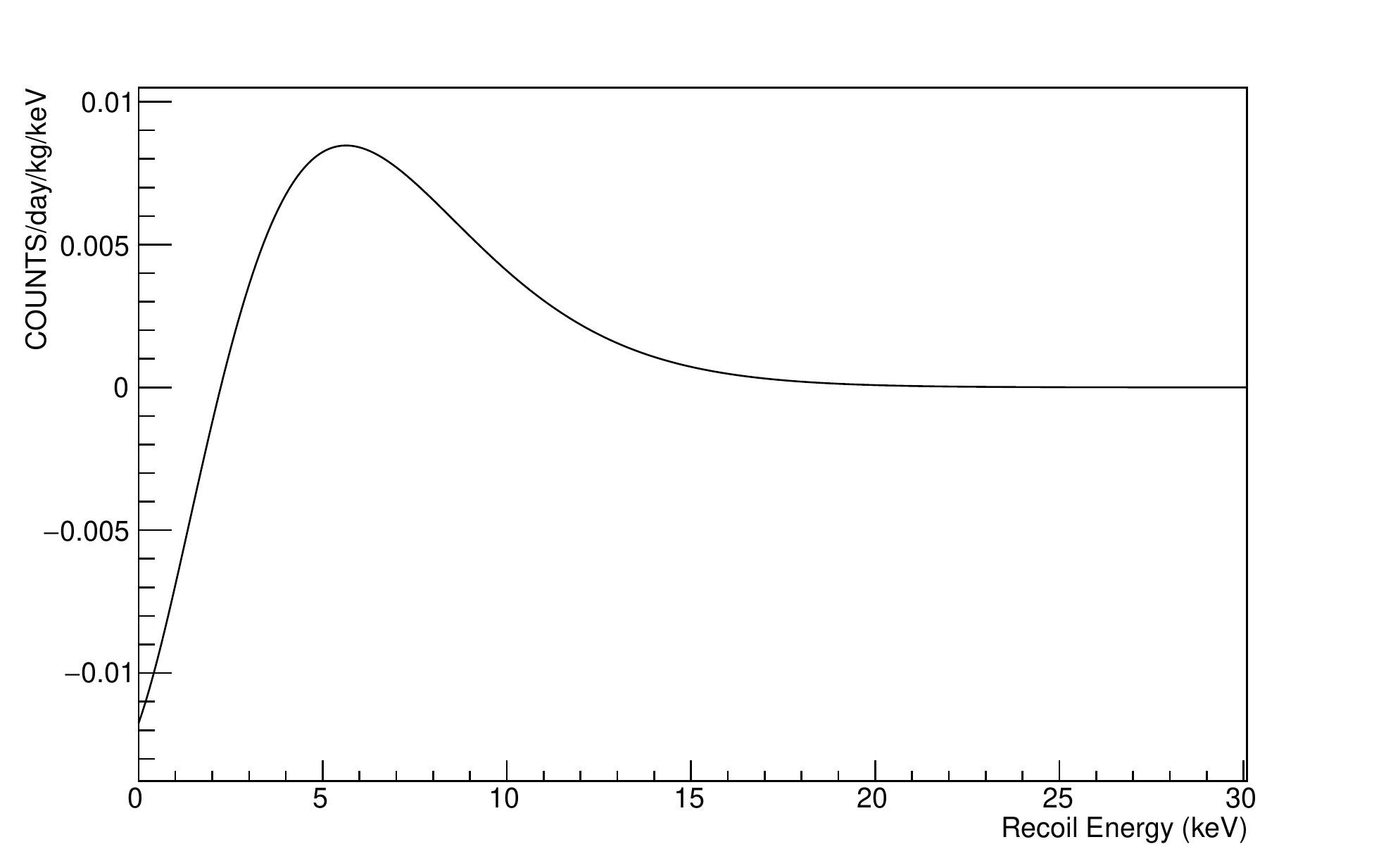}
\caption{The difference of the energy spectra between June and December.}
\label{fg:diff}
\end{wrapfigure}

The annual modulating signal is observed lower than a few tens of keV of the nuclear recoil energy.
The energy spectrum of the recoil nucleus differs between June and December, as shown in Figure \ref{fg:diff}.
In this figure, the target nucleus is sodium, and the horizontal axis is the nuclear recoil energy.
The vertical axis is the expected event rate per keV when the cross section is $1\times10^{-36}$ cm$^{2}$.
The expected difference is as small as about 4\% of the total events of WIMPs. 
Consequently, one needs to construct a large volume detector to obtain sufficient statistical accuracy.
Moreover, the stability of the whole detector system is indispensable; such as temperature, electronic power supply, background conditions, etc.

\subsection{Recent status of WIMPs search}
We need to reduce radioactive contamination both in the detector and in the surrounding materials.
The reduction of contamination in the fiducial volume is the most crucial task since
the detection efficiency of the internal background is almost 100\%,
and it is very difficult to reject by offline analysis.
The present status of the dark matter search is listed in Table \ref{tb:present}.
The background rate in liquid xenon (LXe) detectors is more than three orders of magnitude lower than the NaI(Tl) scintillator.
\begin{table}[ht]
\centering
\caption{Present status of several dark matter search projects.
The BG rate is the typical background rate of electron events in units of \dru .}
\label{tb:present}
\begin{tabular}{l|crrl} \hline
Group & Target & Mass (kg) & BG rate & Ref. \\ \hline
XENON1T & Xe & 1300 & $2\times10^{-4}$ & \cite{Aprile2018} \\
LUX & Xe & 250 & $5\times10^{-4}$ & \cite{Akerib2017} \\
XMASS & Xe & 800 & $4.2\times10^{-3}$ & \cite{Abe2019}\\
DAMIC & Si & 0.036 & 15 & \cite{PhysRevLett.125.241803} \\
SuperCDMS & Ge & 0.6 & 2 & \cite{PhysRevLett.112.041302}\\
ANAIS & NaI & 112.5 & $3.2$ & \cite{Coarasa2019} \\
COSINE-100 & NaI & 106 & $2.7$ & \cite{Adhikari2019} \\
DAMA/LIBRA & NaI & 250 & 1.0 & \cite{Bernabei2008} \\
PICOLON & NaI & 1.25 & $1.5$ & \cite{Kozlov2020} \\ \hline
\end{tabular}
\end{table}

\subsubsection{Liquid Xe detector}
A liquid xenon detector has a significant advantage in low background measurement for dark matter search because of its purity and
event selection power.
The XENON1T group developed an extremely high purity and large volume Xe detector, whose fiducial mass was 1300 kg \cite{Aprile2018}.
The radioactive contamination in the liquid xenon detector consists of natural krypton and
emanated $^{222}$Rn.
Krypton was effectively reduced via cryogenic distillation down to $(0.66\pm0.11)$ ppt \cite{Aprile2018, AprileEpj2018}.

The $^{222}$Rn is hard to remove because it generates various elements, Po, Pb, and Bi, via a sequential decay chain.
Moreover, the $^{222}$Rn is generated from the detector materials and spreads into the fiducial volume.
Nevertheless, they reduced the background by the ionization ratio and the scintillation ratio in each event.
The ratio between scintillation signal $S1$ and ionization signal $S2$ is utilized to discriminate the electron events and nuclear recoil events;
it is called two-phase detector \cite{CLINE2000373, RevModPhys.82.2053, Aprile2018}.
The nuclear recoil event due to WIMPs-nucleus scattering is extracted from a large number of electron background events.

\begin{figure}[htb]
\centering
\includegraphics[width=0.9\linewidth]{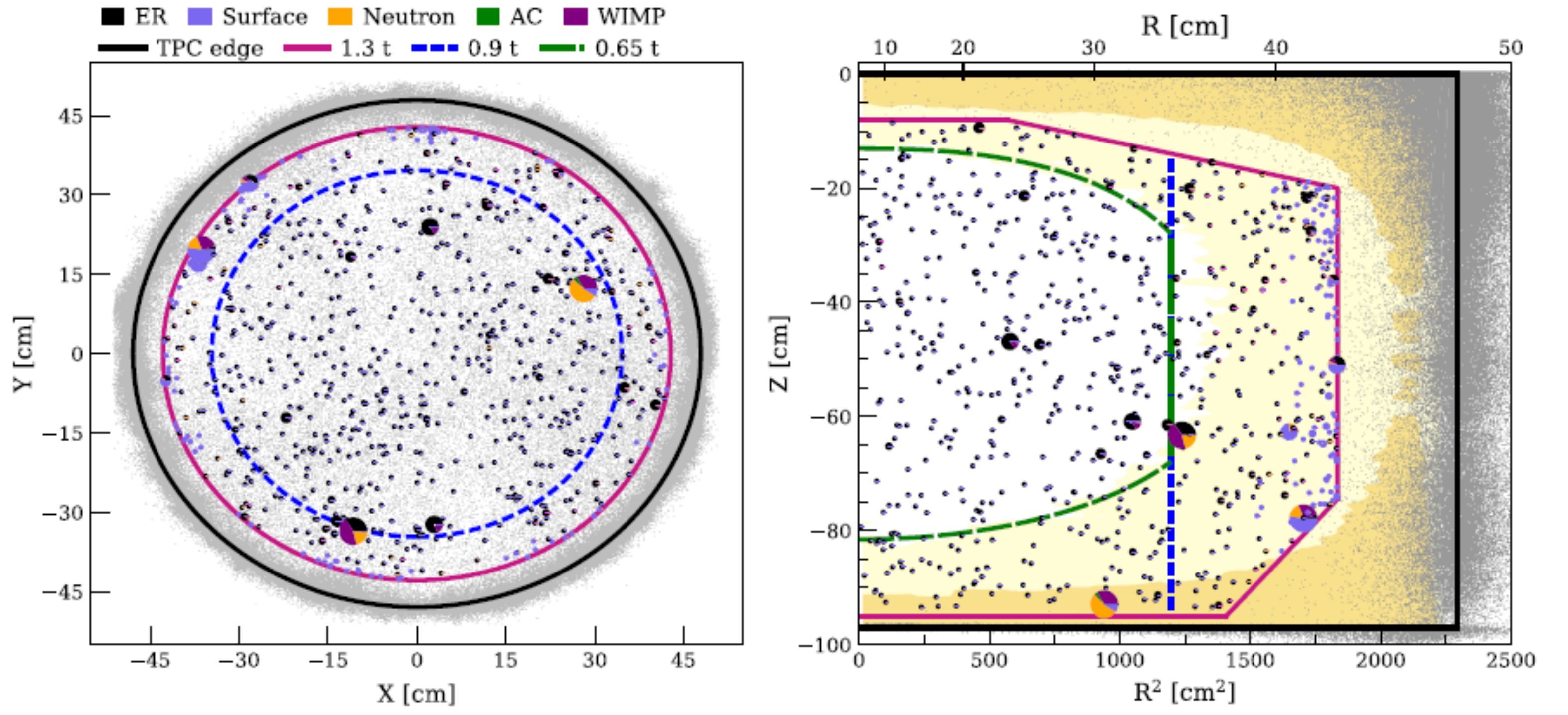}
\caption{Demonstrating of the position selection to reject surface background events \cite{Aprile2018}.}
\label{fg:XENON_pos}
\end{figure}
The information of the event position is helpful to remove the background events.
The position of the events is derived by the ionization signal taken by the time projection chamber (TPC) technique.
Almost all the background events come from the surrounding materials, the detector's housing, photomultiplier tubes (PMT),
and elements of electronic circuits.
These background events interact in the outer region of the fiducial volume (see Figure \ref{fg:XENON_pos}).
The liquid xenon act as the active shield against the background events due to surface contamination.

\begin{wrapfigure}{r}{0.5\linewidth}
\centering
\includegraphics[width=\linewidth]{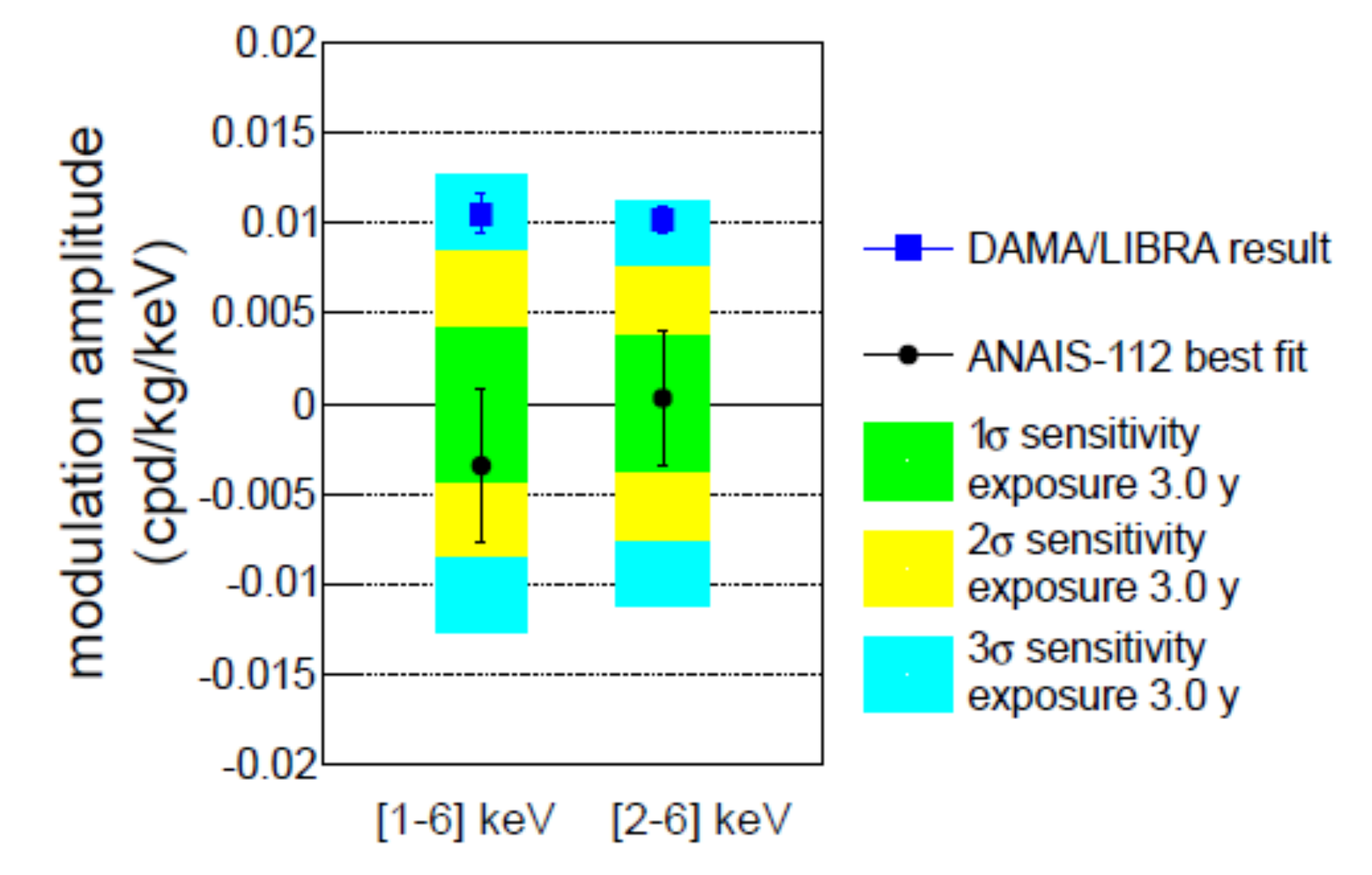}
\caption{The annual modulation amplitude reported ANAIS-112 three years measurement \cite{Amar2021}.
}
\label{fg:ANAIS_3y}
\end{wrapfigure}

\subsubsection{NaI(Tl) detector}
Compared to iodine in the NaI(Tl) detector, xenon differs by one atomic number and has a close mass number, 
so there is little difference for spin-independent interactions with WIMPs.
The first advantage of the NaI(Tl) detector is that it has a high sensitivity to light WIMPs by observing the recoil of light sodium. 
Second, it has a high sensitivity to spin-dependent interactions because of its 100\% odd-numbered nuclei. 
Also, because of the presence of low energy excited states, measuring gamma rays from inelastic scattering is 
advantageous for high sensitivity spin-dependent WIMPs\cite{Ejiri1993}.

The search for an annual modulating signal by a NaI(Tl) scintillator is the present interesting topic in dark matter search.
The DAMA/LIBRA group reported a significant annual modulating signal in the low energy region,
2~keV$_{\mathrm{ee}}\sim 6~$keV$_{\mathrm{ee}}$ \cite{Bernabei2008, Bernabei2018}.
Where keV$_{\mathrm{ee}}$ stands for the observed energy calibrated by the electron energy.

The other groups which apply the NaI(Tl) detector have struggled to reach sufficient sensitivity to test the DAMA/LIBRA's result.
The COSINE group and ANAIS group are continuously searching for the annual modulation signal; however,
they reported no significant modulation in their detectors \cite{Adhikari2019, Amar2021}.
The ANAIS-112 experiment reported no significant modulation signal, which was incompatible with DAMA/LIBRA result
(see Figure \ref{fg:ANAIS_3y}) \cite{Amar2021}.

\subsection{Requirement for the radiation detector for future dark matter search}
Currently, the most sensitive radiation detector for dark matter search consists only xenon
as it is shown in Table \ref{tb:present}.
We need various target nuclei to investigate the property of dark matter candidates.
Consequently, the importance of a highly sensitive radiation detector comparable to the
xenon detectors.

Three reasons for establishing the high sensitivity in xenon detectors are listed below.
\begin{itemize}
\item Sufficiently high-purity of the liquid xenon caused the low background in the fiducial volume.
\item Precise position information enabled the background rejection.
\item Large discrimination power of particles for background rejection.
\end{itemize}
We are developing a solid-state scintillator with the same performance as XENON1T by developing the best
combination of detectors.
The present status of the development to achieve the high performance of the solid-state
scintillator will be described in the following sections.

\section{Development of highly radiopure NaI(Tl) scintillator}
\subsection{Purification of NaI(Tl) crystal}

\begin{wraptable}{r}{0.5\linewidth}
\centering
\caption{The reduction factors for the concentration of Pb ions in the NaI water solution
achieved by the use of various resins.\cite{FushimiPTEP2020}.}
\label{tb:jusi}
\begin{tabular}{cr} \hline
Resin & Reduction factor \\ \hline
A & 1/34 \\
B & 1/64 \\
C & 1/14 \\
D & 1/3 \\ \hline
\end{tabular}
\end{wraptable}
The NaI(Tl) crystal has a large possibility to reduce the intrinsic background.
There is a lot of knowledge for the purification techniques of NaI(Tl) crystal.
The solid-state scintillator is free from contamination after construction because of
its stability.
However, we must be careful not to pollute by radioactive impurity during construction.

We investigated the purification method systematically and found the optimized combination of the methods.
Many groups are developing the reduction methods of NaI(Tl) crystal \cite{FushimiPTEP2020, Park2020, Adhikari2018}.
The present serious radioactive isotopes in our NaI(Tl) crystal are $^{40}$K and $^{210}$Pb.
The potassium in the water solution of NaI forms mainly KI and KOH;
they are as well soluble in water as NaI.\@
We removed potassium ions in the NaI water solution utilizing their significant solubility.

We prepared a saturated NaI water solution at 100 $^{\circ}$C and cooled slowly to room temperature.
We got the pure sediment of NaI by filtration.
The potassium ion remained in the filtrate since the concentration of potassium was sufficiently lower
than its solubility.
The COSINE group reported that the recrystallization method (RC) could remove $^{210}$Pb;
their best result reached down to $10\sim50$ $\mu$Bq/kg \cite{Park2020}.

We considered applying resins to remove lead ions in addition to the recrystallization method.
We have investigated the purification of NaI using a resin. 
We prepared several resins to adsorb heavy ions and compared their effects.
Resins A and B in Table \ref{tb:jusi} are lead ion adsorbing resins manufactured by the same manufacturer. 
Resins C and D in the same table are heavy ion adsorbing resins made by the same manufacturer.
We prepared a NaI water solution that was added 4.8 ppm of Pb ion to test the effectiveness of resins.
The concentration of Pb ion in the processed NaI solution was measured by an inductively coupled plasma mass spectrometry (ICP-MS),
Agilent 7900 in Osaka University and Osaka Sangyo University.
The reduction factors after applying the purification by resins are listed in Table \ref{tb:jusi}.
We decided to use resins A and B after recrystallization.
\begin{table}[ht]
\centering
\caption{The concentration of $^{\mathrm{nat}}$K (ppb),
$^{226}$Ra ($\mu$Bq/kg), $^{232}$Th ($\mu$Bq/kg), and $^{210}$Pb ($\mu$Bq/kg) in
NaI(Tl) scintillators. Characters A to D denotes the resin which is described in Table\ref{tb:jusi}.
RC stands for the recrystallization method.}
\label{tb:junka}
\begin{tabular}{llrrrrc} \hline
Ingot & Method & $^{\mathrm{nat}}$K & $^{226}$Ra & $^{232}$Th & $^{210}$Pb & Ref. \\ \hline
\#24 & A & 2630 & $66\pm11$ & $13\pm8$ & $58\pm26$ & \cite{fushimi2014kamlandpico} \\
\#68 & C+D & 120 & $57\pm7$ & $8.4\pm2.4$ & 7500 & \cite{Kozlov2019} \\
\#71 & RC$\times2$ & $<20$ & $120\pm10$ & $6.8\pm0.8$ & 1500 & \cite{Kozlov2020} \\
\#73 & RC$\times3$ & $<30$ & $44\pm7$ & $7.2\pm0.8$ & 1300 & \cite{Kozlov2020} \\
\#83 & A+B+RC$\times2$ & $<20$ & 11 &22 & 630 & Present work\\
\#85 & A+B+RC$\times2$ & -- & $13\pm4$ & $<3.2$ & $<5.7$ & \cite{FushimiPTEP2020} \\
Our Goal & -- &$<20$ & $<100$ & $<10$ & $<10$ & \\ \hline
\end{tabular}
\end{table}

\begin{wrapfigure}{r}{0.5\linewidth}
\centering
\includegraphics[width=\linewidth]{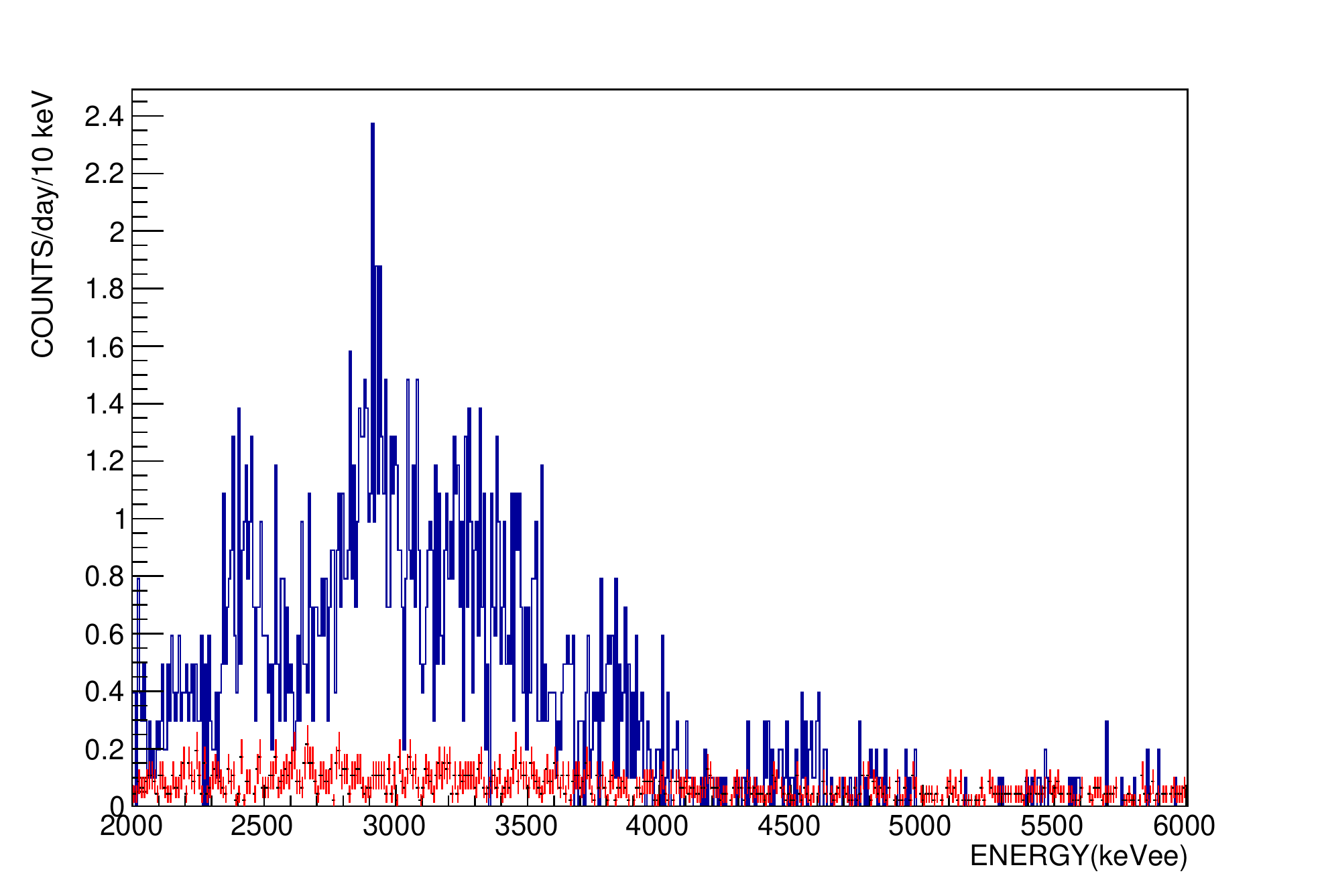}
\caption{The energy spectra of alpha-rays taken by ingots \#24 (Blue) and \#85 (Red) \cite{FushimiPTEP2020}.}
\label{fg:I2485}
\end{wrapfigure}

After making a small NaI(Tl) detector, we measured the radioactive contamination by building a low-background detector system.
The diameter of the NaI(Tl) crystal was 7.62 cm, and the length was 7.62 cm.
We performed various combinations of purification for optimization.
The methods and the results are listed in Table \ref{tb:junka}.
We measured both beta-ray and gamma-ray to determine the concentration of $^{40}$K.
On the other hand,
alpha-ray for uranium series and thorium series isotopes.

We cannot confirm the emitting position of gamma-ray since gamma-rays from electron capture of $^{40}$K is well penetrating in the matter.
We determined the concentration of the origin of $^{40}$K by comparing the beta-ray energy spectrum and the gamma-ray energy spectrum.
We found the recrystallization method effectively removes the $^{40}$K concentration.
We determined to apply double recrystallization for the potassium purification, comparing the double and triple recrystallization results with ingots \#71 and \#73.
We confirmed that the double recrystallization method reduces the potassium concentration
less than 20 ppb \cite{FushimiPTEP2020}.

We measured the alpha-ray intensities of uranium-series and thorium-series isotopes.
The alpha-ray events were extracted by pulse shape discrimination (PSD), identifying the difference of scintillation decay time between
alpha-rays (190 ns) and beta/gamma-rays (230 ns).

We found a significant reduction of alpha-ray intensity in the ingot \#83 as shown in Table \ref{tb:junka}.
We optimized the resin usage condition and successfully derived
a noticeable reduction of alpha-ray intensity in the ingot \#85 as shown in Figure \ref{fg:I2485} \cite{FushimiPTEP2020}.
There was no prominent structure of the alpha-rays emitted by the uranium and the thorium series taken by ingot \#85.
We set an upper limit on the radioactivity of $^{210}$Pb in the ingot \#85 as $5.7$ $\mu$Bq/kg.
We conclude that we found the best combination of the purification methods.

Table \ref{tb:junka} shows the results of our purification processes.
The twice recrystallization is enough to remove potassium ion; however,
it is insufficient to reduce $^{210}$Pb and $^{226}$Ra.
The additional resin usage is practical to remove heavy ions such as lead and radium.
The selection of resin and appropriate use in indispensable to effective purification.
We did not measure potassium concentration in ingot \#85 because we could not install the detector into
a low-background shield in Kamioka underground laboratory.
The reproducibility of potassium reduction was already confirmed by all ingots after \#71.

\section{Test measurement of a large volume NaI(Tl) detector}
\subsection{Design of large volume NaI(Tl)}
We are constructing a large volume NaI(Tl) scintillator array.
The single module is a cylindrical-shaped NaI(Tl) crystal with 12.7 cm diameter and 12.7 cm length.
The crystal was covered with enhanced specular reflector sheet ESR$^{TM}$ provided by 3M to guide the
scintillation photons to optical windows.

The optical windows are attached to the ends of the NaI(Tl) cylinder.
The diameter of the optical window is 7.6 cm, and its thickness is 1.0 cm.
The size of the optical window is fitted to the diameter of a low-background photomultiplier tube,
R11065-20mod provided by Hamamatsu Photonics.

\begin{wrapfigure}{r}{0.5\linewidth}
\includegraphics[width=\linewidth]{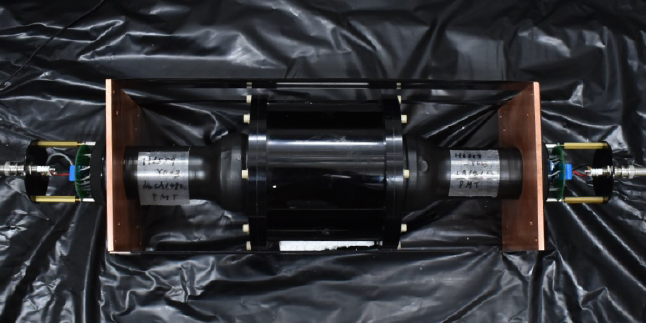}
\caption{A single module of the NaI(Tl) detector.}
\label{fg:largeNaI}
\end{wrapfigure}

We selected black acrylic for a housing material to perform an energy calibration in low energy region.
A periodic energy calibration is one of the essential tasks in the dark matter search
experiment, especially in the low energy region below10 keV$_{\mathrm{ee}}$.
A low-energy X-rays cannot penetrate the detector housing made of copper.
COSINE group performed the energy calibration by K-X ray from electron capture of $^{40}$K,
which is contained in their NaI(Tl) crystals\cite{Park2020}.
It is difficult to apply X-ray from $^{40}$K because of too small event rate;
the expected event rate is less than 1 \dru at the peak.
SABRE group used higher energy gamma ray from $^{241}$Am\cite{Mariani2020}.
A photograph of the prototype detector module is shown in Figure \ref{fg:largeNaI}. 

One cannot keep the NaI(Tl) detector covered with an acrylic housing because moisture is permeable.
However, it is not a problem since the final detector system is contained in an airtight container filled with pure nitrogen gas.
We tested the stability of the NaI(Tl) crystal in an acrylic container whose thickness was 4 mm.
The crystal was kept in a shield filled with pure nitrogen gas; we found no deliquesce or color after one year.

We estimated the effectiveness of low-energy calibration by several popular radioactive sources.
The $^{133}$Ba is one of the suitable radioactive sources which makes sufficiently low energy, 6.4 keV.\@
The energy of K$_{\beta 1}$ X-ray of Cs (a progeny of $^{133}$Ba) is 34.987 keV with 8.4\% intensity \cite{TOI}.
This X-ray is absorbed by the photoelectric effect of the iodine atom followed by the X-ray emission of iodine whose
energy is 28.612 keV (K$_{\alpha 1}$, 46.4\%).
The X-ray of iodine can escape from the NaI(Tl) crystal, and the rest of the energy, 6.4 keV, is observed.

\subsection{Performance of a test module}
We took the energy resolution and energy threshold data by using the prototype detector.
The prototype detector contains a non-purified NaI(Tl) crystal with the exact dimension of the final design.
We attached two photomultiplier tubes (PMT), Hamamatsu R6091, with a 25\% of quantum efficiency.
The trigger of the data acquisition system was generated by a coincidence of both PMTs,
setting each threshold was to get a single photoelectron signal.
This trigger setting is commonly applied to double-readout detector, for example, DAMA/LIBRA
\cite{Bernabei2008}.
Making the trigger with the coincidence signal reduced noise signals from each PMTs
since the rate of the dark current of PMTs is a few kHz.
\begin{wrapfigure}{r}{0.5\linewidth}
\centering
\includegraphics[width=\linewidth]{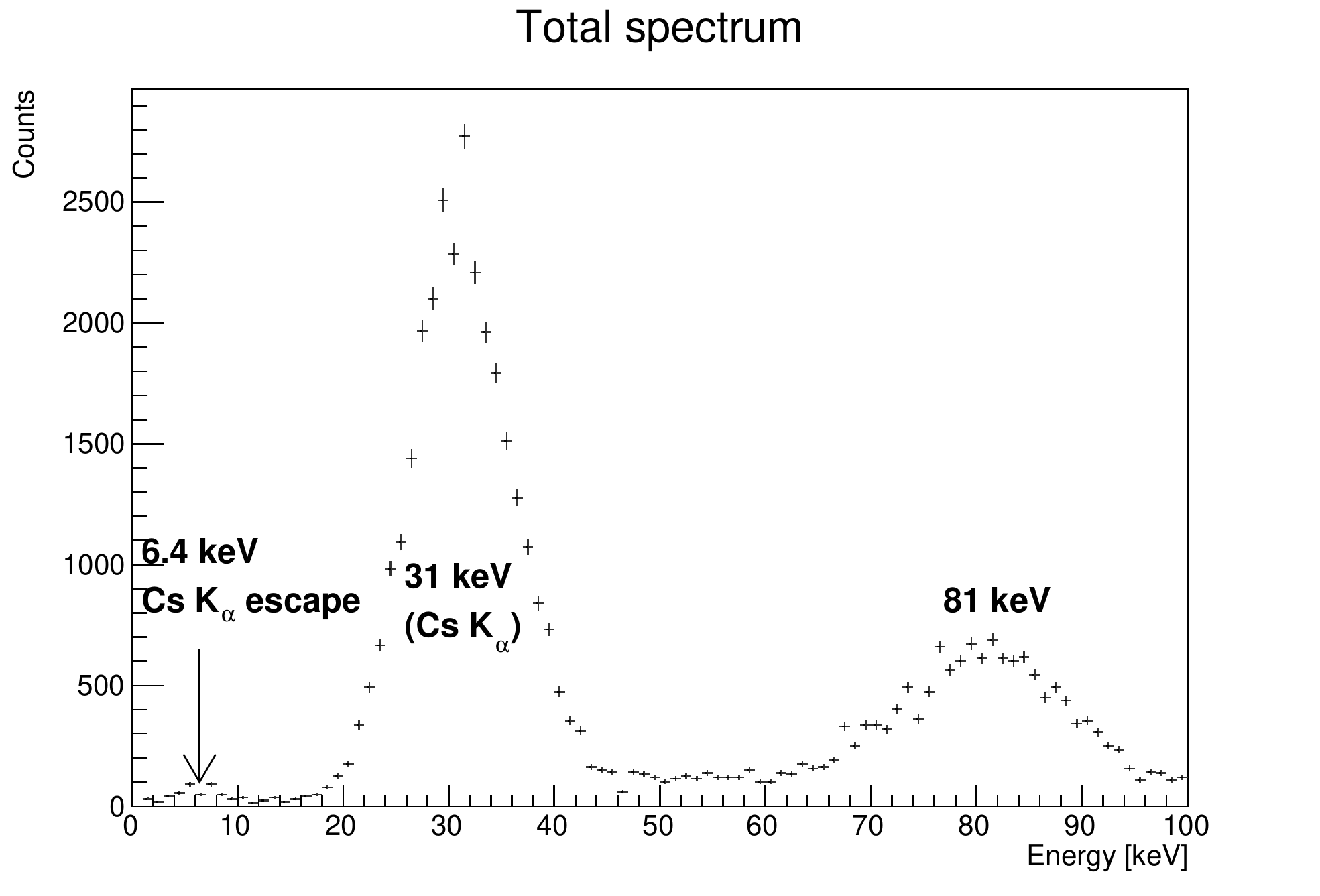}
\caption{The energy spectrum taken by the test module NaI(Tl).
The $^{133}$Ba source was irradiated.}
\label{fg:lowene}
\end{wrapfigure}

We irradiated a $^{133}$Ba source to get pulse shapes of two PMTs.
The pulse shape data was taken by CAMAC data-taking system with flash analog-to-digital-converter
(FADC: REPIC RPC-081).
The pulse shapes from two PMTs were obtained, irradiating a $^{133}$Ba calibration source.
The raw data of the pulse shape contained various noises, pile-up signals, re-triggered pulses and
PMT noises.
We removed each noise by appropriate pulse shape analysis and got an energy spectrum as
shown in Figure \ref{fg:lowene}.

The energy threshold and the energy resolution were 1.6 keV$_{\mathrm{ee}}$ and
26\% at 80 keV$_{\mathrm{ee}}$, respectively.
The performances are due to the low quantum efficiency of PMTs and significant coloring of NaI(Tl)
crystal.
Nevertheless, we confirmed a clear peak of low energy peak around 6.4 keV$_{\mathrm{ee}}$.

The NaI(Tl) scintillator has been reported to have poor linearity of fluorescence in low energy region.
M.Moszynski showed that the degree of non-linearity varies by about 40\% between several keV$_{\mathrm{ee}}$ 
and 100 keV$_{\mathrm{ee}}$ \cite{Moszyski2003}, while L.N.Treflova showed less than 5\% deviation \cite{Trefilova2002}. 
Energy calibration should be performed for each NaI(Tl) to check the non-linearity.
In our test module, the linearity deviation is well within the peak-fitting error between 6.4 keV$_{\mathrm{ee}}$
and 100 keV$_{\mathrm{ee}}$, 
but the linearity deviation was significant for energies below 6.4 keV$_{\mathrm{ee}}$. 
We plan to include $^{40}$K K-X-rays (3 keV) as impurities in the energy calibration for low background measurements.

\section{Prospects}

\begin{wrapfigure}{r}{0.45\linewidth}
\centering
\vspace{-0.5cm}
\includegraphics[width=\linewidth]{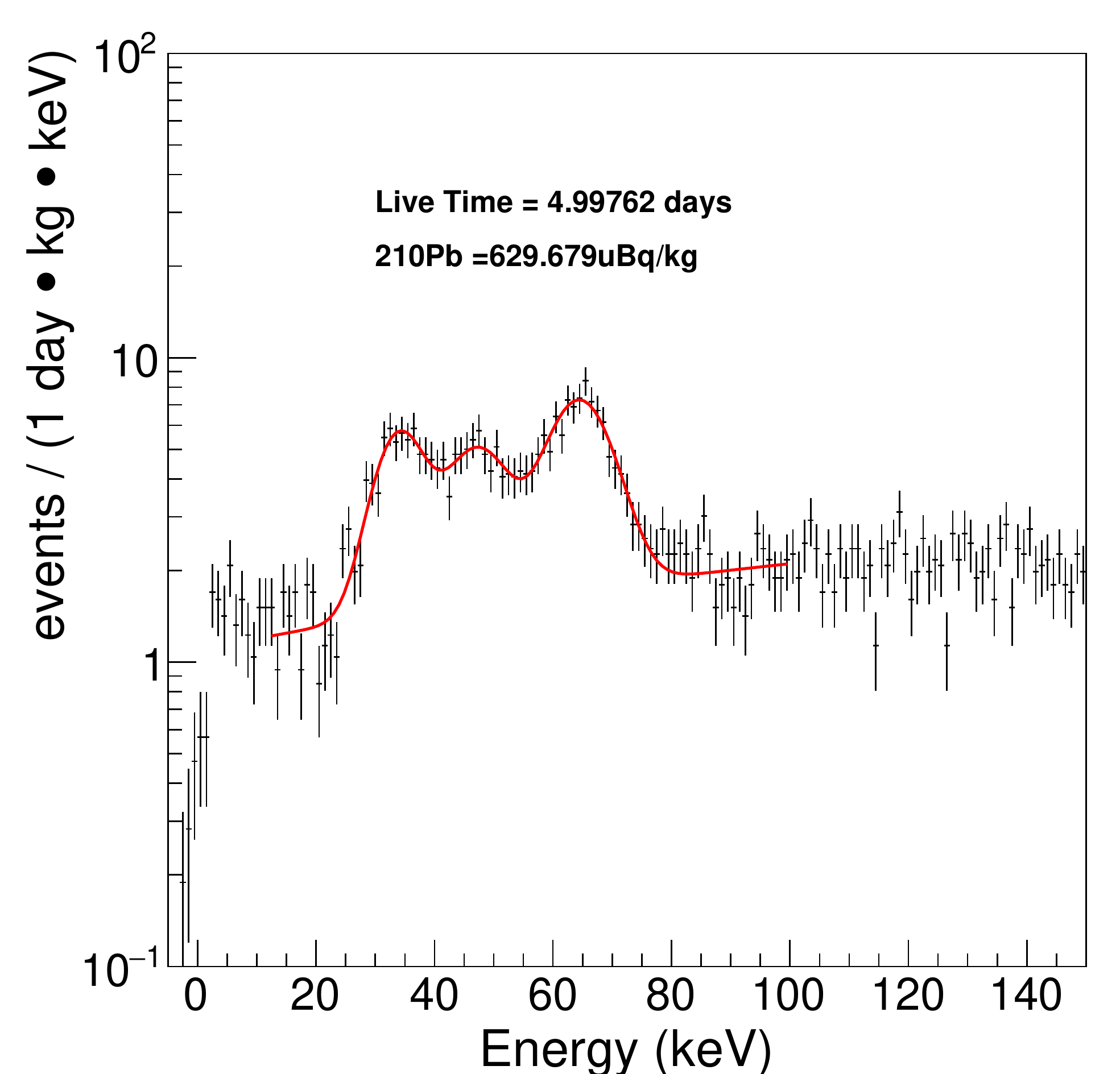}
\caption{The low energy spectrum taken by Ingot\# 83.}
\label{fg:I83}
\end{wrapfigure}

We concluded that $^{210}$Pb mainly contributed to the previous background in our detectors \# 71 to \# 83.
The present low background energy spectrum in Figure \ref{fg:I83} shows three prominent peaks
between 20 \kevee and 60 \kevee .
The lowest and highest peaks are the X-ray and gamma-ray of $^{126}$I ($T_{1/2}=13.11$ day) and $^{125}$I ($T_{1/2}=50.408$ day).
These peaks disappear quickly after installing the NaI(Tl) detector in the underground laboratory.
The central peak at 46.5 \kevee is due to the gamma-ray from $^{210}$Pb.
The concentration of the $^{210}$Pb in the NaI(Tl) crystal was estimated from its peak intensity.
Assuming all the $^{210}$Pb decayed in the NaI(Tl) crystal, the concentration of the $^{210}$Pb is 630 $\mu$Bq/kg.

The event rate near energy threshold was 1.5 \dru as shown in Figure \ref{fg:I83}.
The main origin of this background was $^{210}$Pb in the NaI(Tl) crystal, which was supposed by COSINE-100 experiment.
They estimated the component of their background and the contribution of the $^{210}$Pb was $0.417\pm0.010$ \dru,
in which 0.6 mBq/kg was contained \cite{adhikari2021background}.
They estimated the surface $^{210}$Pb contamination results 0.2 \dru of background.
We expect the background will be less than 1 \dru by future PICOLON experiment.

We will construct the PICOLON phase-I detector, consisting of four NaI(Tl) modules in 2021.
The lower background is expected because of the anti-coincidence measurement between each module.
We already constructed the experimental site in Kamioka underground laboratory at Tohoku University.

The development of a functional detector system is in progress.
The design of the high-selective detector system with NaI(Tl) scintillator is discussed with
KamLAND group.
New functional scintillators with a high performance of particle identification have been developed;
they have a large scintillation output and good performance of particle identification by pulse shape
discrimination \cite{Kamada2017, Iida2020}.

\section{Acknowledgment}
We acknowledge the support of the Kamioka Mining and Smelting Company. This work was
supported by JSPS KAKENHI Grant No. 26104008, 19H00688, 20H05246, and Discretionary expense of the president of Tokushima University.
This work was also supported by the World Premier International Research Center Initiative (WPI Initiative).
We acknowledge Profs.~H.~Sekiya and A.~Takeda of ICRR University of Tokyo, and Prof.~Y.~Takeuchi of Kobe University 
for continuous encouragement and fruitful discussions.

\bibliographystyle{ptephy}
\bibliography{sample}

\end{document}